\documentclass[12pt]{article}
\newcommand{\nc}{\newcommand}
\nc{\fh}{\hat{f}}
\nc{\muh}{\hat{\mu}}
\nc{\nuh}{\hat{\nu}}
\nc{\bib}{\bibitem}
\nc{\al}{\alpha}
\nc{\g}{\gamma}
\nc{\G}{\Gamma}
\nc{\D}{\Delta}
\nc{\eps}{\epsilon}
\nc{\la}{\lambda}
\nc{\La}{\Lambda}
\nc{\var}{\varphi}
\nc{\cg}{{\cal G}}
\nc{\pa}{\partial}
\nc{\nn}{\nonumber \\ }
\nc{\hf}{\frac{1}{2}}  
\nc{\dz}{\frac{dz}{2\pi i}}
\nc{\bin}[2]{\left (\begin{array}{c} {#1}\\ {#2} \end{array}\right )}
\nc{\be}{\begin{equation}}
\nc{\ee}{\end{equation}}
\nc{\bea}{\begin{eqnarray}}
\nc{\eea}{\end{eqnarray}}
\nc{\bra}[1]{\langle {#1}|}
\nc{\ket}[1]{|{#1}\rangle}

\nc{\C}{\mbox{\hspace{1.24mm}\rule{0.2mm}{2.5mm}\hspace{-2.7mm} C}}
\nc{\Nat}{\mbox{\hspace{.04mm}\rule{0.2mm}{2.8mm}\hspace{-1.5mm} N}}

\nc{\HH}{\mbox{\hspace{.04mm}\rule{0.2mm}{2.8mm}\hspace{-1.5mm} H}}

\def\vvdots{\mathinner{\mkern1mu\raise1pt\vbox{\kern7pt\hbox{.}}\mkern2mu
 \raise4pt\hbox{.}\mkern2mu\raise7pt\hbox{.}\mkern1mu}}

\begin{document}

\topmargin -5mm
\oddsidemargin 5mm

\begin{titlepage}
\setcounter{page}{0}

\vspace{8mm}
\begin{center}
{\huge On logarithmic solutions}\\[.4cm]
{\huge to the conformal Ward identities}\\
\vspace{15mm}
{\Large J{\o}rgen Rasmussen}\\[.3cm] 
{\em Department of Mathematics and Statistics, University of Concordia}\\ 
{\em 1455 Maisonneuve W, Montr\'eal, Qu\'ebec, Canada H3G 1M8}\\[.3cm]
rasmusse@crm.umontreal.ca

\end{center}

\vspace{10mm}
\centerline{{\bf{Abstract}}}
\vskip.4cm
\noindent
A general discussion of the conformal Ward identities is presented
in the context of logarithmic conformal field theory with conformal Jordan
cells of rank two. The logarithmic fields are taken to
be quasi-primary. No simplifying assumptions are made about
the operator-product expansions of the primary or logarithmic fields.
Based on a very natural and general ansatz about the form of
the two- and three-point functions, their complete solutions are
worked out. The results are in accordance with and extend
the known results. It is demonstrated, for example, that the correlators
exhibit hierarchical structures similar to the ones found in the literature
pertaining to certain simplifying assumptions.
\end{titlepage}
\newpage
\renewcommand{\thefootnote}{\arabic{footnote}}
\setcounter{footnote}{0}

\section{Introduction}

Logarithmic conformal field theory is essentially based on the appearance of
conformal Jordan cells in the spectrum of fields.
We refer to \cite{Gur}
for the first systematic study of logarithmic conformal field theory, and to
\cite{Flo,Gab,Nic} for recent reviews on the subject. 
The number of fields making up a conformal Jordan cell is called the rank
of the cell. We will focus on conformal Jordan cells of rank two.

We consider the case where the logarithmic fields
in the conformal Jordan cells are quasi-primary, and
discuss the conformal Ward identities which follow.
Without making any simplifying assumptions about the 
operator-product expansions of the fields, we find the general
solutions for two- and three-point functions. Our results
thus cover all the possible cases based on 
primary fields not belonging to conformal
Jordan cells, primary fields belonging to conformal Jordan
cells, and the logarithmic partner fields completing
the conformal Jordan cells. 

We also study the generality of two observations made
under certain simplifying assumptions.
The first observation concerns the expressibility of
the correlators in terms of conformal weights with 
nilpotent parts \cite{MRS}. This is a non-trivial point
as it a priori presumes that the general solutions
to the conformal Ward identities factor accordingly.
We demonstrate that they do.

The second observation concerns a hierarchical structure
for the set of correlators where the links are based on computing derivatives
of the correlators with respect to the conformal weights
\cite{RAK,FloOPE}. Also in this case, we find that the
basic idea extends from the simpler set-up to our general 
situation.

This paper proceeds as follows. After a short introduction
to the conformal Ward identities, we work out the general solutions
for two- and three-point functions. We then affirm the
assertions about conformal weights with nilpotent parts
and the hierarchical structure. We conclude with some
comments on further extensions.

\section{Correlators in logarithmic conformal field theory}

A conformal Jordan cell of rank two consists of two fields:
a primary field, $\Phi$, of conformal weight $\D$ and
its non-primary, `logarithmic' partner field, $\Psi$, 
on which the Virasoro algebra generated by $\{L_n\}$
does not act diagonally. With a conventional relative normalization
of the fields, we have
\bea
 \left[L_n,\Phi(z)\right]&=&\left(z^{n+1}\pa_z+\D(n+1)z^n\right)\Phi(z)\nn
 \left[L_n,\Psi(z)\right]&=&\left(z^{n+1}\pa_z+\D(n+1)z^n\right)\Psi(z)
  +(n+1)z^n\Phi(z)
\label{L}
\eea
It has been suggested by Flohr \cite{Flo-9707} to describe these fields
in a unified way by introducing a nilpotent, yet even, parameter
$\theta$ satisfying $\theta^2=0$. We will follow this idea here,
though use an approach closer to the one employed in \cite{MRS,logsle}.
We thus define the field or unified cell
\be
 \Upsilon(z,\theta)\ =\ \Phi(z)+\theta\Psi(z)
\label{upsilon}
\ee
which is seen to be `primary' of conformal weight $\D+\theta$ as
the commutators (\ref{L}) are replaced by
\be
 \left[L_n,\Upsilon(z,\theta)\right]\ =\ 
  \left(z^{n+1}\pa_z+(\D+\theta)(n+1)z^n\right)\Upsilon(z,\theta)
\label{Lup}
\ee

A primary field belonging to a conformal Jordan cell is referred to
as a `cellular' primary field. 
A primary field {\em not} belonging to a conformal Jordan cell may 
be represented as $\Upsilon(z,0)$, and we will reserve this notation
for these non-cellular primary fields. To avoid ambiguities, we will therefore
refrain from considering unified cells
$\Upsilon(z,\theta)$, as defined in (\ref{upsilon}), for vanishing $\theta$.

\subsection{Conformal Ward identities}

We will consider {\em quasi-primary fields} only, ensuring the projective invariance
of their correlators constructed by sandwiching the fields 
between projectively invariant vacua. 
That is, insertion of any of the three generators
$L_{-1},L_0,L_1$ into a correlator annihilates the correlator.
When expressed in terms of the differential operators (\ref{Lup}),
this is known as the conformal Ward identities which are 
given here for ${\cal N}$-point functions:
\bea
 0&=&\sum_{i=1}^{\cal N}\pa_{z_i}\langle\Upsilon_1(z_1,\theta_1)\dots
  \Upsilon_{\cal N}(z_{\cal N},\theta_{\cal N})\rangle\nn
 0&=&\sum_{i=1}^{\cal N}\left(z_i\pa_{z_i}+\D_i+\theta_i\right)
  \langle\Upsilon_1(z_1,\theta_1)\dots
   \Upsilon_{\cal N}(z_{\cal N},\theta_{\cal N})\rangle\nn
 0&=&\sum_{i=1}^{\cal N}\left(z_i^2\pa_{z_i}+2(\D_i+\theta_i)z_i\right)
  \langle\Upsilon_1(z_1,\theta_1)\dots
   \Upsilon_{\cal N}(z_{\cal N},\theta_{\cal N})\rangle
\label{confward}
\eea
To simplify the notation we introduce the differential
operator
\be
 {\cal L}_1^{\cal N}\ =\ \sum_{i=1}^{\cal N}
  \left(z_i^2\pa_{z_i}+2\D_iz_i\right)
\label{calL}
\ee
in terms of which the third conformal Ward identity reads
\be
 0\ =\ \left({\cal L}_1^{\cal N}+2\sum_{i=1}^{\cal N}\theta_iz_i\right)
   \langle\Upsilon_1(z_1,\theta_1)\dots
   \Upsilon_{\cal N}(z_{\cal N},\theta_{\cal N})\rangle
\label{L1}
\ee
It is noted that a correlator satisfying the first and
third Ward identities (\ref{confward}) automatically
satisfies the second Ward identity. This follows readily
from the commutator $[L_1,L_{-1}]=2L_0$.

The first conformal Ward identity merely imposes
translation invariance on the correlators, allowing
us to express them solely in terms of differences, 
$z_i-z_j$, between the coordinates.

It is stressed that some solutions for
correlators involving non-cellular primary fields $\Upsilon_i(z_i,0)$ 
may be lost if one simply sets the corresponding $\theta_i$
equal to zero in the solutions for non-vanishing $\theta_i$.
This will be illustrated in the following.

Before proceeding, let us indicate how one extracts information
on the individual correlators from solutions to the
conformal Ward identities involving unified cells. In the case of
\be
 \langle\Upsilon_1(z_1,\theta_1)\Upsilon_2(z_2,0)
  \Upsilon_3(z_3,\theta_3)\rangle
\ee 
for example, the identity (\ref{L1}) reads
\be
 0\ =\ \left({\cal L}_1^3+2(\theta_1z_1+\theta_3z_3)\right)
   \langle\Upsilon_1(z_1,\theta_1)\Upsilon_2(z_2,0)
   \Upsilon_3(z_3,\theta_3)\rangle
\ee
A solution to the full set of conformal Ward identities 
is an expression expandable in $\theta_1$
and $\theta_3$. The term proportional to $\theta_1$
but independent of $\theta_3$, for example, 
should then be identified
with $\langle\Psi_1(z_1)\Upsilon_2(z_2,0)\Phi_3(z_3)\rangle$.

By construction, and as illustrated by this example,
correlators involving unified cells and
non-cellular primary fields may thus
be regarded as generating-function correlators whose expansions
in the nilpotent parameters give the individual correlators
involving combinations of cellular primary fields, non-cellular primary 
fields, and logarithmic fields.
Our focus will therefore be on correlators of combinations
of unified cells and non-cellular primary fields. To the best of
our knowledge, most results found in the literature
pertain to correlators involving
unified cells only or non-cellular primary fields only, though
Ref. \cite{FloOPE} does contain a discussion of three-point
functions involving so-called twist fields as examples
of so-called  `pre-logarithmic' fields in the $c=-2$ conformal
field theory.
Those particular results are in accordance with our
general results.
Furthermore, studies of three-point functions involving
unified cells only are most often based
on a simplifying, though physically motivated, assumption
to which we will return in due time.

\subsection{Two-point functions}

We have three situations to analyze, distinguished by the number of
unified cells appearing in the correlator.
The case with non-cellular primary fields only 
is as in ordinary conformal field theory 
and we have the well-known result
\be
 \langle\Upsilon_1(z_1,0)\Upsilon_2(z_2,0)\rangle\ \propto\ 
  \frac{\delta_{\D_1,\D_2}}{z_{12}^{\D_1+\D_2}}
\label{20}
\ee
To simplify the notation, we have introduced the standard
abbreviation $z_{ij}=z_i-z_j$.

We now turn to the situation with at least one unified cell
(i.e., one or two) in the two-point function.
Motivated by the known results for two-point functions
of unified cells only, we consider the following common ansatz
\be
 \langle\Upsilon_1(z_1,\theta_1)\Upsilon_2(z_2,\theta_2)\rangle\ 
  =\ \frac{A(\theta_1,\theta_2)+B(\theta_1,\theta_2)\ln z_{12}}{
  z_{12}^{2h}}
\label{2ans}
\ee
where the dependence of the structure constants $A$ and $B$ on 
$\theta_1$ or $\theta_2$ vanishes if we consider the 
non-cellular primary field $\Upsilon_1(z_1,0)$ or
$\Upsilon_2(z_2,0)$, respectively. The general expansion
of $A$ reads 
\be
 A(\theta_1,\theta_2)\ =\ 
  A^0+A^1\theta_1+A^2\theta_2+A^{12}\theta_1\theta_2
\label{Aexp}
\ee
and similarly for $B$. Imposing (\ref{L1}) results in
\bea
  \langle\Upsilon_1(z_1,\theta_1)\Upsilon_2(z_2,0)\rangle
 &=&\delta_{\D_1,\D_2}\frac{A^1\theta_1}{z_{12}^{\D_1+\D_2}}\nn
  \langle\Upsilon_1(z_1,0)\Upsilon_2(z_2,\theta_2)\rangle
  &=&\delta_{\D_1,\D_2}\frac{A^2\theta_2}{z_{12}^{\D_1+\D_2}}\nn
  \langle\Upsilon_1(z_1,\theta_1)\Upsilon_2(z_2,\theta_2)\rangle
 &=&\delta_{\D_1,\D_2}\frac{A^1\delta_{A^1,A^2}
  \left(\theta_1+\theta_2-2\theta_1\theta_2\ln z_{12}\right)
  +A^{12}\theta_1\theta_2}{z_{12}^{\D_1+\D_2}}
\label{2uni}
\eea
which in terms of individual two-point functions corresponds to
\bea
  \langle\Phi(z_1)\Upsilon(z_2,0)\rangle
  &=&\langle\Upsilon(z_1,0)\Phi(z_2)\rangle
  \ =\  \langle\Phi_1(z_1)\Phi_2(z_2)\rangle
  \ =\ 0\nn
  \langle\Psi(z_1)\Upsilon(z_2,0)\rangle &\propto&
  \frac{\delta_{\D_1,\D_2}}{z_{12}^{\D_1+\D_2}}\nn  
  \langle\Upsilon(z_1,0)\Psi(z_2)\rangle &\propto&
  \frac{\delta_{\D_1,\D_2}}{z_{12}^{\D_1+\D_2}}\nn
 \langle\Phi_1(z_1)\Psi_2(z_2)\rangle
  &=&\langle\Psi_1(z_1)\Phi_2(z_2)\rangle\ =\
  \delta_{\D_1,\D_2}\frac{A^1}{z_{12}^{\D_1+\D_2}}\nn  
 \langle\Psi_1(z_1)\Psi_2(z_2)\rangle&=&\delta_{\D_1,\D_2}
  \frac{A^{12}-2A^1\ln z_{12}}{z_{12}^{\D_1+\D_2}}
\label{2sol}
\eea
Explicit relations similar to the one between $A^1$ and $A^2$ represented
by the delta function in (\ref{2uni}) will be omitted in the following.
As indicated above, the solution (\ref{20})
would have been lost if one were to set $\theta_1=\theta_2=0$
in (\ref{2uni}), whereas the first two solutions in (\ref{2uni})
neatly follow from the last solution in (\ref{2uni})
if one sets $\theta_2=0$ or $\theta_1=0$, respectively.

\subsection{Three-point functions}

We now have four situations to analyze, again
characterized by the number of
unified cells appearing in the correlator.
As for two-point functions, the case with non-cellular primary 
fields only is as in ordinary conformal field theory 
and we have the well-known result
\be
 \langle\Upsilon_1(z_1,0)\Upsilon_2(z_2,0)\Upsilon_3(z_3,0)
  \rangle\ \propto\ 
  \frac{1}{z_{12}^{\D_1+\D_2-\D_3}z_{23}^{-\D_1+\D_2+\D_3}
   z_{13}^{\D_1-\D_2+\D_3}}
\label{30}
\ee

For the combined three-point functions, associativity and the
results on two-point functions suggest that we consider
the following ansatz 
\bea
 &&\langle\Upsilon_1(z_1,\theta_1)\Upsilon_2(z_2,\theta_2)
  \Upsilon_3(z_3,\theta_3)\rangle\nn
 &=&
  \{A(\theta_1,\theta_2,\theta_3)+
   B_{1}(\theta_1,\theta_2,\theta_3)\ln z_{12}
   +B_{2}(\theta_1,\theta_2,\theta_3)\ln z_{23}
   +B_{3}(\theta_1,\theta_2,\theta_3)\ln z_{13}\nn
 &&\ +D_{11}(\theta_1,\theta_2,\theta_3)\ln^2 z_{12}
  +D_{12}(\theta_1,\theta_2,\theta_3)\ln z_{12}\ln z_{23}
  +D_{13}(\theta_1,\theta_2,\theta_3)\ln z_{12}\ln z_{13}\nn
  &&\ +D_{22}(\theta_1,\theta_2,\theta_3)\ln^2 z_{23}
  +D_{23}(\theta_1,\theta_2,\theta_3)\ln z_{23}\ln z_{13}
  +D_{33}(\theta_1,\theta_2,\theta_3)\ln^2 z_{13}\}\nn
 &\times&\left(z_{12}^{-h_{1}}z_{23}^{-h_{2}}z_{13}^{-h_{3}}\right)
\label{3ans}
\eea
Here $h_{i}$ is $\theta$-independent while
\be
 A(\theta_1,\theta_2,\theta_3)\ =\ 
  A^0+A^1\theta_1+A^2\theta_2+A^3\theta_3
  +A^{12}\theta_1\theta_2+A^{23}\theta_2\theta_3+A^{13}\theta_1\theta_3
  +A^{123}\theta_1\theta_2\theta_3
\label{Atheta}
\ee
and similarly for $B_{i}$ and $D_{ij}$.
Imposing the Ward identities (i.e., on this ansatz,
(\ref{L1}) suffices), corresponds
to the following conditions, obtained from considering the part independent 
of logarithms
\bea
 0&=&(2\D_1-h_1-h_3+2\theta_1)A(\theta_1,\theta_2,\theta_3)
  +B_1(\theta_1,\theta_2,\theta_3)+B_3(\theta_1,\theta_2,\theta_3)\nn
 0&=&(2\D_2-h_1-h_2+2\theta_2)A(\theta_1,\theta_2,\theta_3)
  +B_1(\theta_1,\theta_2,\theta_3)+B_2(\theta_1,\theta_2,\theta_3)\nn 
 0&=&(2\D_3-h_2-h_3+2\theta_3)A(\theta_1,\theta_2,\theta_3)
  +B_2(\theta_1,\theta_2,\theta_3)+B_3(\theta_1,\theta_2,\theta_3)
\label{3con0}
\eea
the part linear in logarithms
\bea
 0&=&(2\D_1-h_1-h_3+2\theta_1)B_1(\theta_1,\theta_2,\theta_3)
  +2D_{11}(\theta_1,\theta_2,\theta_3)+D_{13}(\theta_1,\theta_2,\theta_3)\nn
 0&=&(2\D_2-h_1-h_2+2\theta_2)B_1(\theta_1,\theta_2,\theta_3)
  +2D_{11}(\theta_1,\theta_2,\theta_3)+D_{12}(\theta_1,\theta_2,\theta_3)\nn 
 0&=&(2\D_3-h_2-h_3+2\theta_3)B_1(\theta_1,\theta_2,\theta_3)
  +D_{12}(\theta_1,\theta_2,\theta_3)+D_{13}(\theta_1,\theta_2,\theta_3)\nn
 0&=&(2\D_1-h_1-h_3+2\theta_1)B_2(\theta_1,\theta_2,\theta_3)
  +D_{12}(\theta_1,\theta_2,\theta_3)+D_{23}(\theta_1,\theta_2,\theta_3)\nn
 0&=&(2\D_2-h_1-h_2+2\theta_2)B_2(\theta_1,\theta_2,\theta_3)
  +D_{12}(\theta_1,\theta_2,\theta_3)+2D_{22}(\theta_1,\theta_2,\theta_3)\nn 
 0&=&(2\D_3-h_2-h_3+2\theta_3)B_2(\theta_1,\theta_2,\theta_3)
  +2D_{22}(\theta_1,\theta_2,\theta_3)+D_{23}(\theta_1,\theta_2,\theta_3)\nn
 0&=&(2\D_1-h_1-h_3+2\theta_1)B_3(\theta_1,\theta_2,\theta_3)
  +D_{13}(\theta_1,\theta_2,\theta_3)+2D_{33}(\theta_1,\theta_2,\theta_3)\nn
 0&=&(2\D_2-h_1-h_2+2\theta_2)B_3(\theta_1,\theta_2,\theta_3)
  +D_{13}(\theta_1,\theta_2,\theta_3)+D_{23}(\theta_1,\theta_2,\theta_3)\nn 
 0&=&(2\D_3-h_2-h_3+2\theta_3)B_3(\theta_1,\theta_2,\theta_3)
  +D_{23}(\theta_1,\theta_2,\theta_3)+2D_{33}(\theta_1,\theta_2,\theta_3)
\label{3con1}
\eea
and the part quadratic in logarithms
\bea
 0&=&(2\D_1-h_1-h_3+2\theta_1)D_{ij}(\theta_1,\theta_2,\theta_3),
  \ \ \ \ \ \ \ \ \ 1\leq i\leq j\leq3\nn
 0&=&(2\D_2-h_1-h_2+2\theta_2)D_{ij}(\theta_1,\theta_2,\theta_3),
  \ \ \ \ \ \ \ \ \ 1\leq i\leq j\leq3\nn
 0&=&(2\D_3-h_2-h_3+2\theta_3)D_{ij}(\theta_1,\theta_2,\theta_3),
  \ \ \ \ \ \ \ \ \ 1\leq i\leq j\leq3
\label{3con2}
\eea
These apply whether or not the individual $\theta$s vanish,
even if $\theta_1=\theta_2=\theta_3=0$ as in (\ref{30}).
In the further analysis, one should distinguish between the different
numbers
of unified cells, that is, the numbers of non-vanishing $\theta$s.
Also, it is understood that an $A^1$, for example, appearing in the study
of one set of correlators (related through one or several 
Jordan-cell structures)
a priori is independent of an $A^1$ appearing in a different set (not related
to the former through a Jordan-cell structure).

Now, it is not hard to show that we in every case have
\be
 h_1\ =\ \D_1+\D_2-\D_3,\ \ \ \ h_2\ =\ -\D_1+\D_2+\D_3,\ \ \ \ h_3\ =\
  \D_1-\D_2+\D_3
\label{h}
\ee
meaning that these identities apply to all combinations of vanishing
or non-vanishing $\theta$s. In the case where 
$\theta_1=\theta_2=\theta_3=0$, there is only one solution to
the conditions (\ref{3con0}-\ref{3con2}) and one recovers 
(\ref{30}) with $A^0$ as the proportionality constant.

In the case where $\theta_1\neq0$ while $\theta_2=\theta_3=0$,
we find
\be
  \langle\Upsilon_1(z_1,\theta_1)\Upsilon_2(z_2,0)\Upsilon_3(z_3,0)
  \rangle\ =\ 
  \frac{A^0+A^1\theta_1+A^0\theta_1(-\ln z_{12}+\ln z_{23}-\ln z_{13})}{
   z_{12}^{\D_1+\D_2-\D_3}z_{23}^{-\D_1+\D_2+\D_3}
   z_{13}^{\D_1-\D_2+\D_3}}
\label{31}
\ee
which in terms of the individual correlators reads
\bea
  \langle\Phi_1(z_1)\Upsilon_2(z_2,0)\Upsilon_3(z_3,0)
  \rangle&=&\frac{A^0}{z_{12}^{\D_1+\D_2-\D_3}z_{23}^{-\D_1+\D_2+\D_3}
   z_{13}^{\D_1-\D_2+\D_3}}\nn
 \langle\Psi_1(z_1)\Upsilon_2(z_2,0)\Upsilon_3(z_3,0)
  \rangle&=&\frac{A^1-A^0\ln\frac{z_{12}z_{13}}{z_{23}}}{
   z_{12}^{\D_1+\D_2-\D_3}z_{23}^{-\D_1+\D_2+\D_3}
   z_{13}^{\D_1-\D_2+\D_3}}
\label{31ind}
\eea
The other two cases with only one unified cell are treated similarly
and the corresponding correlators may be obtained from 
(\ref{31}) and (\ref{31ind}) by appropriately permuting the indices.
We note that there in each case are two a priori
independent structure constants.
Before commenting on the structure of these results, let us complete
the analysis of the conditions (\ref{3con0}-\ref{3con2}).

In the case where $\theta_1,\theta_2\neq0$ while $\theta_3=0$,
we find
\bea
  &&\langle\Upsilon_1(z_1,\theta_1)\Upsilon_2(z_2,\theta_2)\Upsilon_3(z_3,0)
  \rangle\nn 
  &=&\left\{A^0+A^1\theta_1+A^2\theta_2+A^{12}\theta_1\theta_2
  +\left(-A^0\theta_1-A^0\theta_2-(A^1+A^2)\theta_1\theta_2\right)\ln z_{12}
    \right.\nn
  &+&\left(A^0\theta_1-A^0\theta_2+(-A^1+A^2)\theta_1\theta_2\right)\ln z_{23}
   +\left(-A^0\theta_1+A^0\theta_2+(A^1-A^2)\theta_1\theta_2\right)\ln z_{13}\nn
   &+&\left. A^0\theta_1\theta_2\left(\ln^2z_{12}-\ln^2z_{23}-\ln^2z_{13}
   +2\ln z_{23}\ln z_{13}\right)\right\}\nn
  &\times&z_{12}^{-\D_1-\D_2+\D_3}z_{23}^{\D_1-\D_2-\D_3}
   z_{13}^{-\D_1+\D_2-\D_3}
\label{32}
\eea
which in terms of the individual correlators reads
\bea
  \langle\Phi_1(z_1)\Phi_2(z_2)\Upsilon_3(z_3,0)
  \rangle&=&\frac{A^0}{z_{12}^{\D_1+\D_2-\D_3}z_{23}^{-\D_1+\D_2+\D_3}
   z_{13}^{\D_1-\D_2+\D_3}}\nn
 \langle\Psi_1(z_1)\Phi_2(z_2)\Upsilon_3(z_3,0)
  \rangle&=&\frac{A^1-A^0\ln\frac{z_{12}z_{13}}{z_{23}}}{
   z_{12}^{\D_1+\D_2-\D_3}z_{23}^{-\D_1+\D_2+\D_3}
   z_{13}^{\D_1-\D_2+\D_3}}\nn
 \langle\Phi_1(z_1)\Psi_2(z_2)\Upsilon_3(z_3,0)
  \rangle&=&\frac{A^2-A^0\ln\frac{z_{12}z_{23}}{z_{13}}}{
   z_{12}^{\D_1+\D_2-\D_3}z_{23}^{-\D_1+\D_2+\D_3}
   z_{13}^{\D_1-\D_2+\D_3}}\nn
 \langle\Psi_1(z_1)\Psi_2(z_2)\Upsilon_3(z_3,0)
  \rangle&=&\frac{A^{12}-A^1\ln\frac{z_{12}z_{23}}{z_{13}}-
   A^2\ln\frac{z_{12}z_{13}}{z_{23}}+A^0\ln\frac{z_{12}z_{23}}{z_{13}}
    \ln\frac{z_{12}z_{13}}{z_{23}}}{z_{12}^{\D_1+\D_2-\D_3}
    z_{23}^{-\D_1+\D_2+\D_3}z_{13}^{\D_1-\D_2+\D_3}}
   \label{32ind}
\eea
The other two cases with two unified cells are treated similarly
and the corresponding correlators may be obtained from 
(\ref{32}) and (\ref{32ind}) by an appropriate permutation of the indices.
We note that there in each case
are four a priori independent structure constants.

In the case with three unified cells, that is, $\theta_1,\theta_2,\theta_3\neq0$,
we find
\bea
  &&\langle\Upsilon_1(z_1,\theta_1)\Upsilon_2(z_2,\theta_2)
    \Upsilon_3(z_3,\theta_3)\rangle\nn 
  &=&\{A^1\theta_1+A^2\theta_2+A^3\theta_3+A^{12}\theta_1\theta_2
   +A^{23}\theta_2\theta_3+A^{13}\theta_1\theta_3
   +A^{123}\theta_1\theta_2\theta_3\nn
 &+&\left((-A^1-A^2)\theta_1\theta_2+(A^2-A^3)\theta_2\theta_3
   +(A^1-A^3)\theta_1\theta_3\right.\nn
  &&\left. +(A^{12}-A^{23}-A^{13})\theta_1\theta_2\theta_3\right)\ln z_{12}\nn
&+&\left((-A^1+A^2)\theta_1\theta_2+(-A^2-A^3)\theta_2\theta_3
   +(-A^1+A^3)\theta_1\theta_3\right.\nn
  &&\left. +(-A^{12}+A^{23}-A^{13})\theta_1\theta_2\theta_3\right)\ln z_{23}\nn
&+&\left((A^1-A^2)\theta_1\theta_2+(-A^2+A^3)\theta_2\theta_3
   +(-A^1-A^3)\theta_1\theta_3\right.\nn
  &&\left. +(-A^{12}-A^{23}+A^{13})\theta_1\theta_2\theta_3\right)\ln z_{13}\nn
 &+&(-A^1-A^2+A^3)\theta_1\theta_2\theta_3\ln^2z_{12}
  +2A^2\theta_1\theta_2\theta_3\ln z_{12}\ln z_{23}\nn
 &+&2A^1\theta_1\theta_2\theta_3\ln z_{12}\ln z_{13}
  +(A^1-A^2-A^3)\theta_1\theta_2\theta_3\ln^2z_{23}\nn
  &+&2A^3\theta_1\theta_2\theta_3\ln z_{23}\ln z_{13}
  +(-A^1+A^2-A^3)\theta_1\theta_2\theta_3\ln^2z_{13}\}\nn
 &\times&z_{12}^{-\D_1-\D_2+\D_3}z_{23}^{\D_1-\D_2-\D_3}
   z_{13}^{-\D_1+\D_2-\D_3}
\label{33}
\eea
which in terms of the individual correlators reads
\bea
  \langle\Phi_1(z_1)\Phi_2(z_2)\Phi_3(z_3)
  \rangle&=&0\nn
 \langle\Psi_1(z_1)\Phi_2(z_2)\Phi_3(z_3)
  \rangle&=&\frac{A^1}{z_{12}^{\D_1+\D_2-\D_3}
    z_{23}^{-\D_1+\D_2+\D_3}z_{13}^{\D_1-\D_2+\D_3}}\nn
 \langle\Phi_1(z_1)\Psi_2(z_2)\Phi_3(z_3)
  \rangle&=&\frac{A^2}{z_{12}^{\D_1+\D_2-\D_3}
    z_{23}^{-\D_1+\D_2+\D_3}z_{13}^{\D_1-\D_2+\D_3}}\nn
 \langle\Phi_1(z_1)\Phi_2(z_2)\Psi_3(z_3)
  \rangle&=&\frac{A^3}{z_{12}^{\D_1+\D_2-\D_3}
    z_{23}^{-\D_1+\D_2+\D_3}z_{13}^{\D_1-\D_2+\D_3}}\nn
 \langle\Psi_1(z_1)\Psi_2(z_2)\Phi_3(z_3)
  \rangle&=&\frac{A^{12}-A^1\ln\frac{z_{12}z_{23}}{z_{13}}
   -A^2\ln\frac{z_{12}z_{13}}{z_{23}}}{
  z_{12}^{\D_1+\D_2-\D_3}
    z_{23}^{-\D_1+\D_2+\D_3}z_{13}^{\D_1-\D_2+\D_3}}\nn
 \langle\Phi_1(z_1)\Psi_2(z_2)\Psi_3(z_3)
  \rangle&=&\frac{A^{23}-A^2\ln\frac{z_{23}z_{13}}{z_{12}}
   -A^3\ln\frac{z_{12}z_{23}}{z_{13}}}{
  z_{12}^{\D_1+\D_2-\D_3}
    z_{23}^{-\D_1+\D_2+\D_3}z_{13}^{\D_1-\D_2+\D_3}}\nn
 \langle\Psi_1(z_1)\Phi_2(z_2)\Psi_3(z_3)
  \rangle&=&\frac{A^{13}-A^1\ln\frac{z_{23}z_{13}}{z_{12}}
   -A^3\ln\frac{z_{12}z_{13}}{z_{23}}}{
  z_{12}^{\D_1+\D_2-\D_3}
    z_{23}^{-\D_1+\D_2+\D_3}z_{13}^{\D_1-\D_2+\D_3}}\nn
 \langle\Psi_1(z_1)\Psi_2(z_2)\Psi_3(z_3)
  \rangle&=&\left(A^{123}-A^{12}\ln\frac{z_{23}z_{13}}{z_{12}}
  -A^{23}\ln\frac{z_{12}z_{13}}{z_{23}}
  -A^{13}\ln\frac{z_{12}z_{23}}{z_{13}}\right.\nn
  &&+A^1\ln\frac{z_{12}z_{23}}{z_{13}}\ln\frac{z_{23}z_{13}}{z_{12}}
   +A^2\ln\frac{z_{12}z_{13}}{z_{23}}\ln\frac{z_{23}z_{13}}{z_{12}}\nn
  &&\left. +A^3\ln\frac{z_{12}z_{23}}{z_{13}}\ln\frac{z_{12}z_{13}}{z_{23}}
   \right)z_{12}^{-\D_1-\D_2+\D_3}z_{23}^{\D_1-\D_2-\D_3}
   z_{13}^{-\D_1+\D_2-\D_3} \nn
\label{33ind}
\eea
We note that there are seven a priori independent structure
constants. In the literature, on the other hand, one deals with 
three structure constants only (see \cite{Flo}, for example). 
This discrepancy is due to 
an assumption usually made in available studies of 
three-point functions. It concerns a particular property
of the cellular primary fields which we will address presently.

Primary fields are called {\em proper primary} 
if their operator-product expansions
with each other cannot produce a logarithmic field.
It is argued in \cite{FloCluster} (see also \cite{GG}) that correlators 
not involving improper primary fields satisfy
\bea
 \langle\Psi_1(z_1)\Phi_2(z_2)\dots\Phi_{\cal N}(z_{\cal N})\rangle
 &=& \langle\Phi_1(z_1)\Psi_2(z_2)
  \Phi_3(z_3)\dots\Phi_{\cal N}(z_{\cal N})\rangle\nn
 &\vdots&\nn
 &=&\langle\Phi_1(z_1)\dots\Phi_{{\cal N}-1}(z_{{\cal N}-1})
  \Psi_{\cal N}(z_{\cal N})\rangle
\label{cluster}
\eea
in particular, and that the general form of the individual
three-point functions of logarithmic fields and cellular primary fields
hence read
\bea
  \langle\Phi_1(z_1)\Phi_2(z_2)\Phi_3(z_3)
  \rangle&=&0\nn
 \langle\Psi_1(z_1)\Phi_2(z_2)\Phi_3(z_3)
  \rangle&=& \langle\Phi_1(z_1)\Psi_2(z_2)\Phi_3(z_3)
  \rangle\ =\ \langle\Phi_1(z_1)\Phi_2(z_2)\Psi_3(z_3)
  \rangle\nn
  &=&\frac{C_{123;1}}{z_{12}^{\D_1+\D_2-\D_3}
    z_{23}^{-\D_1+\D_2+\D_3}z_{13}^{\D_1-\D_2+\D_3}}\nn
 \langle\Psi_1(z_1)\Psi_2(z_2)\Phi_3(z_3)
  \rangle&=&\frac{C_{123;2}-2C_{123;1}\ln z_{12}}{
  z_{12}^{\D_1+\D_2-\D_3}
    z_{23}^{-\D_1+\D_2+\D_3}z_{13}^{\D_1-\D_2+\D_3}}\nn
 \langle\Phi_1(z_1)\Psi_2(z_2)\Psi_3(z_3)
  \rangle&=&\frac{C_{123;2}-2C_{123;1}\ln z_{23}}{
  z_{12}^{\D_1+\D_2-\D_3}
    z_{23}^{-\D_1+\D_2+\D_3}z_{13}^{\D_1-\D_2+\D_3}}\nn
 \langle\Psi_1(z_1)\Phi_2(z_2)\Psi_3(z_3)
  \rangle&=&\frac{C_{123;2}-2C_{123;1}\ln z_{13}}{
  z_{12}^{\D_1+\D_2-\D_3}
    z_{23}^{-\D_1+\D_2+\D_3}z_{13}^{\D_1-\D_2+\D_3}}\nn
 \langle\Psi_1(z_1)\Psi_2(z_2)\Psi_3(z_3)
  \rangle&=&\{C_{123;3}-C_{123;2}(\ln z_{12}+\ln z_{23}+\ln z_{13})
     \nn
 &&+C_{123;1}(2\ln z_{12}\ln z_{23}+2\ln z_{12}\ln z_{13}
    +2\ln z_{23}\ln z_{13}\nn
 &&-\ln^2z_{12}-\ln^2z_{23}-\ln^2z_{13})\}\nn
 &\times&z_{12}^{-\D_1-\D_2+\D_3}z_{23}^{\D_1-\D_2-\D_3}
   z_{13}^{-\D_1+\D_2-\D_3} 
\label{3cluster}
\eea
We may recover this result from (\ref{33ind}) 
by relating the structure constants appearing there as
\be
 A^1\ =\ A^2\ =\ A^3,\ \ \ \ \ \ \ A^{12}\ =\ A^{23}\ =\ A^{13}
\label{clustercond}
\ee
in which case the correlators
(\ref{33ind}) are seen to reduce to (\ref{3cluster}) with
$C_{123;1}=A^1$, $C_{123;2}=A^{12}$, and $C_{123;3}=A^{123}$.

Regarding the reduction in the number of unified cells
in a three-point function, it is observed that setting $\theta_3=0$
in (\ref{33}) does not reproduce the full expression (\ref{32}) but
only the part  
independent of $A^0$. Setting $\theta_2=0$ in (\ref{32})
or $\theta_1=0$ in (\ref{31}), on the other hand, neatly reproduces
the expressions (\ref{31}) and (\ref{30}), respectively.

According to the general results above, a logarithmic singularity may
appear in a three-point function involving only one
logarithmic field as long as at least one of the other two
(primary) fields is non-cellular. This is in contrast to the
situation based on conformal Jordan cells only,
where at least two logarithmic fields are required to
have a logarithmic singularity. Likewise, a singularity
quadratic in logarithms may appear in a three-point function
with two logarithmic fields and one non-cellular primary
field, while such a singularity cannot appear if
the primary field is cellular.

\subsection{In terms of weights with nilpotent parts}

It has been discussed how the correlators of unified cells only may
be represented compactly if one considers the nilpotent 
parameter $\theta_i$ as part of a generalized conformal weight
given by $\Delta_i+\theta_i$ \cite{MRS}. 
A general version of this assertion is of course very natural from
the point of view of the extended Virasoro action (\ref{Lup}).
It nevertheless {\em presumes} that the {\em general} solution to the
conformal Ward identities may be factored accordingly.
This has been shown to be the case when the simplifying assumption
about the cellular primary fields being proper primary 
is imposed. The extension to our general set-up is discussed
in the following and is found to affirm the assertion.

The two-point functions may thus be represented as
\bea
  \langle\Upsilon_1(z_1,0)\Upsilon_2(z_2,0)\rangle
 &=&\delta_{\D_1,\D_2}\frac{A^0}{z_{12}^{\D_1+\D_2}}\nn
   \langle\Upsilon_1(z_1,\theta_1)\Upsilon_2(z_2,0)\rangle
 &=&\delta_{\D_1,\D_2}\frac{A^1\theta_1}{z_{12}^{(\D_1+\theta_1)+\D_2}}\nn
   \langle\Upsilon_1(z_1,\theta_1)\Upsilon_2(z_2,\theta_2)\rangle
 &=&\delta_{\D_1,\D_2}\frac{A^1(\theta_1+\theta_2)
  +A^{12}\theta_1\theta_2}{z_{12}^{(\D_1+\theta_1)+(\D_2+\theta_2)}}
\label{2unith}
\eea
The similar expression for the
correlator $\langle\Upsilon_1(z_1,0)\Upsilon_2(z_2,\theta_2)\rangle$
is obtained from the second one by interchanging the indices.

It is straightforward to verify that 
the three-point functions may be represented as
\bea
 \langle\Upsilon_1(z_1,0)\Upsilon_2(z_2,0)\Upsilon_3(z_3,0)
  \rangle&=& 
   \frac{A^0}{z_{12}^{\D_1+\D_2-\D_3}z_{23}^{-\D_1+\D_2+\D_3}
    z_{13}^{\D_1-\D_2+\D_3}}\nn
  \langle\Upsilon_1(z_1,\theta_1)\Upsilon_2(z_2,0)\Upsilon_3(z_3,0)
  \rangle&=& 
  \frac{A^0+A^1\theta_1}{
   z_{12}^{(\D_1+\theta_1)+\D_2-\D_3}z_{23}^{-(\D_1+\theta_1)+\D_2+\D_3}
   z_{13}^{(\D_1+\theta_1)-\D_2+\D_3}}
\label{301th}
\eea
and
\bea
  &&\langle\Upsilon_1(z_1,\theta_1)\Upsilon_2(z_2,\theta_2)\Upsilon_3(z_3,0)
  \rangle\nn 
  &&\ \ \ \ =\ \frac{A^0+A^1\theta_1+A^2\theta_2+A^{12}\theta_1\theta_2}{
   z_{12}^{(\D_1+\theta_1)+(\D_2+\theta_2)-\D_3}
   z_{23}^{-(\D_1+\theta_1)+(\D_2+\theta_2)+\D_3}
   z_{13}^{(\D_1+\theta_1)-(\D_2+\theta_2)+\D_3}} \nn
 &&\langle\Upsilon_1(z_1,\theta_1)\Upsilon_2(z_2,\theta_2)
    \Upsilon_3(z_3,\theta_3)\rangle\nn 
 &&\ \ \ \ =\ \frac{A^1\theta_1+A^2\theta_2+A^3\theta_3
   +A^{12}\theta_1\theta_2
   +A^{23}\theta_2\theta_3+A^{13}\theta_1\theta_3
   +A^{123}\theta_1\theta_2\theta_3}{
   z_{12}^{(\D_1+\theta_1)+(\D_2+\theta_2)-(\D_3+\theta_3)}
   z_{23}^{-(\D_1+\theta_1)+(\D_2+\theta_2)+(\D_3+\theta_3)}
   z_{13}^{(\D_1+\theta_1)-(\D_2+\theta_2)+(\D_3+\theta_3)}} 
\label{323th}
\eea
The remaining four combinations are obtained by appropriate
permutations in the indices.

As already indicated, it is not clear a priori that the general solutions 
to the conformal Ward identities (\ref{confward}) based on the 
ans\"atze (\ref{2ans}) and (\ref{3ans}) reduce to expressions
which may be factored as in (\ref{2unith}), (\ref{301th}) and (\ref{323th}).
Our analysis has demonstrated that this is indeed the case.

\subsection{Derivatives with respect to the conformal weights}

Acting on either
\be
 W_2\ =\ \frac{\delta_{\D_1,\D_2}}{z_{12}^{\D_1+\D_2}}
\label{W2}
\ee
or
\be
 W_3\ =\ \frac{1}{z_{12}^{\D_1+\D_2-\D_3}z_{23}^{-\D_1+\D_2+\D_3}
  z_{13}^{\D_1-\D_2+\D_3}}
\label{W3}
\ee
we may substitute derivatives with respect to the conformal weights
by multiplicative factors according to
\be
 \pa_{\D_1}\ =\ \pa_{\D_2}\ \rightarrow\ -2\ln z_{12}
\label{Dln2}
\ee
or 
\be
 \pa_{\D_1}\ \rightarrow\ -\ln\frac{z_{12}{z_{13}}}{z_{23}},\ \ \ \ \ \ \ 
 \pa_{\D_2}\ \rightarrow\ -\ln\frac{z_{12}{z_{23}}}{z_{13}},\ \ \ \ \ \ \ 
 \pa_{\D_3}\ \rightarrow\ -\ln\frac{z_{23}{z_{13}}}{z_{12}}
\label{Dln3}
\ee
respectively. This simple observation allows us to represent
the correlators involving logarithmic fields as follows:
\bea
  \langle\Psi_1(z_1)\Upsilon_2(z_2,0)\rangle&=&A^1W_2\nn
  \langle\Psi_1(z_1)\Phi_2(z_2)\rangle&=&A^1W_2\nn
  \langle\Psi_1(z_1)\Psi_2(z_2)\rangle&=&\left(A^{12}+A^2\pa_{\D_1}
    +A^1\pa_{\D_2}\right)W_2\nn
   \langle\Psi_1(z_1)\Upsilon_2(z_2,0)
   \Upsilon_3(z_3,0)\rangle&=&\left(A^1+A^0\pa_{\D_1}\right)W_3\nn
 \langle\Psi_1(z_1)\Phi_2(z_2)
   \Upsilon_3(z_3,0)\rangle&=&\left(A^1+A^0\pa_{\D_1}\right)W_3\nn
 \langle\Psi_1(z_1)\Psi_2(z_2)
   \Upsilon_3(z_3,0)\rangle&=&\left(A^{12}+A^1\pa_{\D_2}
    +A^2\pa_{\D_1}+A^0\pa_{\D_1}\pa_{\D_2}\right)W_3\nn
   \langle\Psi_1(z_1)\Phi_2(z_2)
   \Phi_3(z_3)\rangle&=&A^1W_3\nn
 \langle\Psi_1(z_1)\Psi_2(z_2)
   \Phi_3(z_3)\rangle&=&\left(A^{12}+A^2\pa_{\D_1}
    +A^1\pa_{\D_2}\right)W_3\nn
 \langle\Psi_1(z_1)\Psi_2(z_2)
   \Psi_3(z_3)\rangle&=&
  \left(A^{123}+A^{23}\pa_{\D_1}+A^{13}\pa_{\D_2}
   +A^{12}\pa_{\D_3}\right.\nn
  &&+\left.A^3\pa_{\D_1}\pa_{\D_2}+A^1\pa_{\D_2}\pa_{\D_3}
   +A^2\pa_{\D_1}\pa_{\D_3}\right)W_3
\label{3D}
\eea 
in addition to expressions obtained by appropriately permuting
the indices.
One may therefore represent the correlators hierarchically as
\bea
 \langle\Psi_1(z_1)\Upsilon_2(z_2,0)\rangle&=&A^1W_2
   +\pa_{\D_1}\langle\Phi_1(z_1)\Upsilon_2(z_2,0)\rangle\nn
  \langle\Psi_1(z_1)\Phi_2(z_2)\rangle&=&A^1W_2
    +\pa_{\D_1}\langle\Phi_1(z_1)\Phi_2(z_2)\rangle\nn
  \langle\Psi_1(z_1)\Psi_2(z_2)\rangle&=&A^{12}W_2
   +\pa_{\D_1}\langle\Phi_1(z_1)\Psi_2(z_2)\rangle
    +\pa_{\D_2}\langle\Psi_1(z_1)\Phi_2(z_2)\rangle\nn
  &-&\pa_{\D_1}\pa_{\D_2}\langle\Phi_1(z_1)\Phi_2(z_2)\rangle
\label{2h}
\eea
in the case of two-point functions, and
\bea
   \langle\Psi_1(z_1)\Upsilon_2(z_2,0)
   \Upsilon_3(z_3,0)\rangle&=&A^1W_3
   +\pa_{\D_1}\langle\Phi_1(z_1)\Upsilon_2(z_2,0)
   \Upsilon_3(z_3,0)\rangle\nn
 \langle\Psi_1(z_1)\Phi_2(z_2)
   \Upsilon_3(z_3,0)\rangle&=&A^1W_3
   +\pa_{\D_1}\langle\Phi_1(z_1)\Phi_2(z_2)
   \Upsilon_3(z_3,0)\rangle\nn
 \langle\Psi_1(z_1)\Psi_2(z_2)
   \Upsilon_3(z_3,0)\rangle&=&A^{12}W_3
    +\pa_{\D_1}\langle\Phi_1(z_1)\Psi_2(z_2)
      \Upsilon_3(z_3,0)\rangle\nn
  &+&\pa_{\D_2}\langle\Psi_1(z_1)\Phi_2(z_2)
      \Upsilon_3(z_3,0)\rangle\nn
  &-&\pa_{\D_1}\pa_{\D_2}\langle\Phi_1(z_1)\Phi_2(z_2)
    \Upsilon_3(z_3,0)\rangle\nn
   \langle\Psi_1(z_1)\Phi_2(z_2)
   \Phi_3(z_3)\rangle&=&A^1W_3
   +\pa_{\D_1}\langle\Phi_1(z_1)\Phi_2(z_2)
   \Phi_3(z_3)\rangle\nn
 \langle\Psi_1(z_1)\Psi_2(z_2)
   \Phi_3(z_3)\rangle&=&A^{12}W_3
    +\pa_{\D_1}\langle\Phi_1(z_1)\Psi_2(z_2)
      \Phi_3(z_3)\rangle\nn
  &+&\pa_{\D_2}\langle\Psi_1(z_1)\Phi_2(z_2)
      \Phi_3(z_3)\rangle
  -\pa_{\D_1}\pa_{\D_2}\langle\Phi_1(z_1)\Phi_2(z_2)
    \Phi_3(z_3)\rangle\nn
 \langle\Psi_1(z_1)\Psi_2(z_2)
   \Psi_3(z_3)\rangle&=&A^{123}W_3
   +\pa_{\D_1}\langle\Phi_1(z_1)\Psi_2(z_2)
   \Psi_3(z_3)\rangle\nn
  &+&\pa_{\D_2}\langle\Psi_1(z_1)\Phi_2(z_2)
   \Psi_3(z_3)\rangle
  +\pa_{\D_3}\langle\Psi_1(z_1)\Psi_2(z_2)
   \Phi_3(z_3)\rangle\nn
  &-&\pa_{\D_1}\pa_{\D_2}\langle\Phi_1(z_1)\Phi_2(z_2)
   \Psi_3(z_3)\rangle\nn
  &-&\pa_{\D_2}\pa_{\D_3}\langle\Psi_1(z_1)\Phi_2(z_2)
   \Phi_3(z_3)\rangle\nn
  &-&\pa_{\D_1}\pa_{\D_3}\langle\Phi_1(z_1)\Psi_2(z_2)
   \Phi_3(z_3)\rangle\nn
  &+&\pa_{\D_1}\pa_{\D_2}\pa_{\D_3}\langle\Phi_1(z_1)\Phi_2(z_2)
   \Phi_3(z_3)\rangle
\label{3h}
\eea
in the case of three-point functions. As above, the remaining correlators 
may be obtained by appropriately permuting the indices.
Similar results in the particular case of proper primary fields (\ref{3cluster})
have already appeared in the literature \cite{FloOPE}, see also \cite{RAK}.

We finally wish to re-address the conformal Ward identities in the realm
of these hierarchical structures. Since the latter are the same in all
the cases, we will focus on the most complex scenario, the one
involving the three-point function of three logarithmic fields.
The conformal Ward identity following from inserting $L_1$ into
such a correlator may be written
\be
 0\ =\ \left({\cal L}_1^3+2z_1\delta_1+2z_2\delta_2+2z_3\delta_3\right)
   \langle\Psi_1(z_1)\Psi_2(z_2)\Psi_3(z_3)\rangle
\label{L1Psi}
\ee
As is conventional in logarithmic conformal field theory, we have introduced
here the operator $\delta_i$ acting (in the case of a conformal Jordan cell
of rank two) on the fields in a correlator as
\be
 \delta_i\Psi_j(z_j)\ =\ \delta_{ij}\Phi_j(z_j),\ \ \ \ \ \ \ \ \ \ 
 \delta_i\Phi_j(z_j)\ =\ 0
\label{di}
\ee
in addition to $ \delta_i\Upsilon_j(z_j,0)=0$.
This means that the conformal Ward identity (\ref{L1Psi}) reads
\bea
 0&=&{\cal L}_1^3\langle\Psi_1(z_1)\Psi_2(z_2)\Psi_3(z_3)\rangle
  +2z_1\langle\Phi_1(z_1)\Psi_2(z_2)\Psi_3(z_3)\rangle\nn
 &+&2z_2\langle\Psi_1(z_1)\Phi_2(z_2)\Psi_3(z_3)\rangle
  +2z_3\langle\Psi_1(z_1)\Psi_2(z_2)\Phi_3(z_3)\rangle
\label{L1Psi2}
\eea
This condition is easily verified using (\ref{3D}). It is stressed, though,
that it is only with hindsight that these structures appear natural.

\section{Conclusion}

We have studied the conformal Ward identities for quasi-primary
fields appearing in logarithmic conformal field theory based on  
conformal Jordan cells of rank two.  
Even though our results are based on an ansatz, it
appears natural to suspect that they constitute the general solution
for two- and three-point functions.

We anticipate that one, in a straightforward manner, may extract 
general information about the operator-product expansions
underlying the correlators we have found. 
This is an interesting enterprise we intend to undertake.

As already mentioned, our results
pertain to conformal Jordan cells of rank two. We hope
to study the case of general rank elsewhere. Partial
results in this direction may be found in \cite{RAK,FloOPE}.
Conformal Jordan cells of infinite rank have been introduced
in \cite{infcell}.

We have found that the results presented in this paper may
be extended to affine Jordan cells appearing in certain
logarithmic extensions of Wess-Zumino-Witten models
\cite{affinecell}. The general solutions in these models also satisfy
the Knizhnik-Zamolodchikov equations and are found to reduce,
by hamiltonian reduction, to
the solutions provided in the present paper. 

Another natural extension of the present work which would
be interesting to pursue, is the general solution to the superconformal
Ward identities appearing in logarithmic superconformal
field theory. Results in this direction may be found in 
\cite{KAG}.
\vskip.5cm
\noindent{\em Acknowledgements}
\vskip.1cm
\noindent  The author thanks Michael Flohr for very helpful comments.

\end{document}